\documentclass[pre,aps,amsmath,eqsecnum,unsortedaddress,nofootinbib,preprint]{revtex4}

\usepackage{amssymb}
\usepackage{amsmath}
\usepackage{graphicx}
\usepackage{amsbsy}
\usepackage{color}

\begin{document}
\title{Competition for hydrogen bond formation in the helix-coil transition and protein folding.}

\author{A.V. Badasyan}
\email{abadasyan@gmail.com}
\affiliation{Dipartimento di Scienze Molecolari e Nanosistemi, Universit\`a Ca' Foscari Venezia,
Calle Larga S. Marta DD2137, I-30123 Venezia, Italy}

\author{Sh.A. Tonoyan}
\affiliation{Department of Molecular Physics, Yerevan State University,\\ A.Manougian Str.1, 375025, Yerevan, Armenia}

\author{Y.Sh.Mamasakhlisov}
\affiliation{Department of Molecular Physics, Yerevan State University,\\ A.Manougian Str.1, 375025, Yerevan, Armenia}

\author{Achille Giacometti}

\affiliation{Dipartimento di Scienze Molecolari e Nanosistemi, Universit\`a Ca' Foscari Venezia,
Calle Larga S. Marta DD2137, I-30123 Venezia, Italy}

\author{A.S.Benight}

\affiliation{Departments of Chemistry and Physics, Portland State University,\\ 1719 S.W. 10th Ave., Portland, OR 97207-0751, USA}

\author{V.F.Morozov}

\affiliation{Department of Molecular Physics, Yerevan State University,\\ A.Manougian Str.1, 375025, Yerevan, Armenia}

\date{\today}

\begin{abstract}

The problem of the helix-coil transition of biopolymers in explicit solvents, like water, with the ability for hydrogen bonding with solvent is addressed analytically using a suitably modified version of the Generalized Model of Polypeptide Chains. Besides the regular helix-coil transition, an additional coil-helix or reentrant transition is also found at lower temperatures. The reentrant transition arises due to competition between polymer-polymer and polymer-water hydrogen bonds. The balance between the two types of hydrogen bonding can be shifted to either direction through changes not only in temperature, but also by pressure, mechanical force, osmotic stress or other external influences. Both polypeptides and polynucleotides are considered within a unified formalism. Our approach provides an explanation of the experimental difficulty of observing the reentrant transition with pressure; and underscores the advantage of pulling experiments for studies of DNA. Results are discussed and compared with those reported in a number of recent publications with which a significant level of agreement is obtained.
 
\end{abstract}


\maketitle
\section{Introduction}
\label{sec:introduction}
The helix-coil transition is a central event in many genetic processes of living matter \cite{cantor,molbiol}. A number of different approaches to describing the helix-coil transition in biopolymers have appeared in the literature; many of them based on spin models \cite{flory,gros,mattice,goldstein,bad10,stanprot,biopoly1,biopoly2,physa,ananik}. As shown in Refs.~\onlinecite{bad10,biopoly1,biopoly2}, the frequently used Zimm-Bragg \cite{ZB} and Lifson-Roig \cite{LR} models are particular cases of a more general model based on a Potts-like Hamiltonian. Within these spin models the influence of solvent on the helix-coil transition has been the topic of many studies \cite{polsher}. This is because biopolymers inside cells act in the presence of water with dissolved metal ions and other low and high molecular weight compounds \cite{molbiol,cantor}. Many of the approaches to describing solvent interactions are quite diverse and often intractable analytically.

Solvents differ by the mechanisms of their action on polymer chains dissolved in them. For instance, water is a natural solvent for biopolymers and in addition to forming hydrogen bonded networks in the bulk, water also has the ability to form hydrogen bonds with biopolymers. Water is the most important biological solvent with very interesting properties due to the large number of anomalies present in the water phase diagram that still are a matter of debate \cite{pnas10}. 

 With a coupled Ising-Potts model Vause, Walker and Goldstein \cite{VWG} achieved significant success describing lower critical solution points in hydrogen-bonded mixtures. Accounting for these studies, during the last few decades slightly different approaches for describing the water phase diagram were introduced by Stokely, Debenedetti, Stanley and others \cite{stanley_group}. Within a Bell-Lavis spin model, the possibility of a reentrant phase diagram between low and high density phases of water was recently pointed out  \cite{fiore09}. 

The key success in these studies was due to the proper accounting for the fact that hydrogen bonding between two species takes place only at special orientations of water molecules, reflecting the tetrahedral symmetry of water systems. In principle, applications of spin models allows capture of the most important features of solvents with the ability for H-bond formation, i.e. the directional character of hydrogen bonds and the large entropic changes that occur due to this directionality. Especially useful in this sense is the Potts model. 

In view of the above-mentioned facts, for description of conformational transitions in biopolymers it seems natural to exploit spin models for both the polymer and solvent. Within this context, a detailed description of the water anomalies is often irrelevant for two reasons. First, the most interesting and biologically important events in biopolymers (helix-coil transition, protein folding) occur under conditions, where water is in the bulk liquid state far from freezing or other critical points. Second, it is widely accepted that polymer hydration is short-ranged in a direction normal to the polymer axis, and polymer-water interactions affect only one (two, in some cases) layer(s) of the water network \cite{saenger,bloomfield}. Conversely, it is very important for consideration of polymers to properly account for the possibility of water-polymer hydrogen bond formation.

Within a spin description Goldstein attempted to include both polymer and simplified water models \cite{goldstein}. His approach included different Potts variables for descriptions of states of repeated units of polymer and solvent molecules. He showed for any ratio of the two energies, provided that the energy of polymer-solvent hydrogen bonds is larger than the polymer-polymer hydrogen bond energy, the possibility of the helix-coil transition at high temperatures, and the reentrant coil-helix transition at low temperatures.

The helix-coil transition is a constituent part of protein folding and is closely related to cold denaturation as well. Currently there is no agreement on a general molecular mechanism that results in both cold and heat denaturation in proteins. Accordingly, there is no generally accepted approach or model that allows descriptions of both the direct and reentrant transition on the same footing. 

A rather interesting attempt linking the changes in secondary and tertiary structures of polymers was reported in Ref.~\cite{walk}. It was shown that reentrant isotropic-nematic phase transitions can be mediated by helix-coil transformations within individual liquid crystal molecules. Reentrance of the isotropic phase was shown to be driven by the inverted helix-coil transformation. It is well-known that the coil-globule transition in polymers strongly depends on the rigidity of the polymer chain \cite{gros}. Therefore it is not surprising that by altering chain rigidity via the helix-coil transition it is possible to tune the coil-globule equilibrium. We reach the conclusion that the reentrant helix-coil transition, arising from interactions with water through the reentrant rigidity effect, could be the origin of reentrant folding and cold denaturation. 

In the framework of a zipper model, Hansen and Bakk (HB) have recently accounted for competing effects between protein and water that lead to cold denaturation \cite{hansen,bakk}. Another attempt of describing the influence of water on folding was reported by the authors of Ref.~\onlinecite{stanprot} who included directional features of water hydrogen bonds using Potts variables and modeling folding as Go-like. They showed the possibility of cold denaturation depended on the ratio of water-water and water-protein hydrogen bond energies. It was concluded that effects of pressure on water density are key to understanding cold denaturation in proteins. The reported phase diagram agrees quite well with the experimental data obtained on bovine pancreatic ribonuclease A. Among the vast arsenal of possibilities, the authors of Ref.~\onlinecite{stanprot} used one of the simplest models of water \cite{stanley_group}, and still were unable to analytically average over solvent degrees of freedom. 

De Los Rios and Caldarelli \cite{caldarelli} used a similar description for water to account for cold denaturation in proteins. They borrowed some simplifications from the HB protein folding model and concluded that the effective attractive interaction between hydrophobic species should depend on temperature. Near the end of their paper they cautioned that care should be exercised when defining generally valid effective interaction potentials among amino-acids, since during folding the protein exists in an ever changing conformational environment and hence strongly depends on the amino-acid distribution and on interactions with water.

Within the assumption that the transition free energy of a protein, i.e. the free-energy difference between the native and denatured states, is a quadratic function of pressure and temperature, Hawley \emph{et al} constructed a melting theory \cite{hawley_theory}. Based on their analysis the resulting bell-shaped 3D curve ($\Delta G(P,T)$) was sliced at zero surface to obtain an elliptical phase diagram of the transition between the native and denatured states. Quite good agreement was achieved with some of the published experimental results (see \cite{smeller,tanaka} for review and comparison with experiments). The same (Hawley) theory was also applied to describe the DNA reentrant melting transition \cite{dubins}.

In a rather different context, to produce both cold and heat denaturations, Riccio, Ascolese and Graziano \cite{graziano} discussed importance of considering the free energy difference between the transition states as a quadratic function of temperature. With the same goal, the authors of Ref.~\onlinecite{chan} inserted a temperature-dependent hydrophobic attraction as a curve with a maximum, without justifying the origin of such dependence.

Polypeptides and proteins are identical from the viewpoint of their interactions with water. The same interactions must also affect secondary structure formation in DNA. Although there are large structural differences between polypeptides and polynucleotides both macromolecules share some important similarities, the most relevant ones being the stabilization of helical structure by hydrogen bonds and having water as the most important (native) solvent \footnote{Not surprising since both macromolecules are constituent parts of the genome system of a cell and function in the same aqueous environment. For example, the H-bonds between DNA strands are locally broken to access the genetic information and the protein is synthesized according to the genetic code read in the same environment.}. Generally, when water-DNA interactions are considered the primary focus is on screening negative charges of DNA phosphates by water and dissolved counterions. Whereas much less attention is usually paid to formation of water-DNA hydrogen bonds. For many years the reentrant transition in DNA was elusive and not observed experimentally \cite{nordmeier}.  More recently experimental evidence for the reentrant melting has been obtained through high pressure measurements \cite{dubins,macgregor} and single-molecule pulling experiments \cite{rouzina,williams}. Our approach provides the theoretical basis from which the reported experimental observations of pressure induced denaturation in DNA \cite{macgregor} can be interpreted and provides an explanation of why the reentrant melting transition of DNA is observed in experiments at much higher pressure values than is the case of proteins.

While it is clear that competition between polymer-polymer and water-polymer hydrogen bonds has to be taken into account, the microscopic origin and consequences of such competition remains unclear. Analysis of this point stimulated the investigation reported here. The aim of this paper is to join current theoretical approaches for characterizing conformational transitions of biopolymers in solvent environments where the possibility of hydrogen bonding with solvent is explicitly considered. To this aim, results of previous studies are built upon and the influence of solvent interactions on the helix-coil transition is considered within a particular Potts-like model \cite{biopoly1}. We show how orientational degrees of freedom of water can be summed out analytically resulting in an effective Hamiltonian term with a temperature-dependent interaction energy. This re-normalized  temperature dependent energy is related to the free energy of the melting transition, and the phase diagram for the helix-coil transition is reported. Our approach provides a description of temperature, pressure, pH, osmotic pressure and denaturant effects on an equal footing, thus providing a convenient framework within which to investigate rather complex situations. 
\section{The GMPC model with solvent}
For this work consider the Generalized Model of Polypeptide Chains (GMPC) \cite{biopoly1,biopoly2,biopoly2004,mplb2005,ictp,hetero,arsen} with the
following Hamiltonian,
\begin{equation}
\label{hamtotal}
\begin{gathered}
   - \beta H_{\text{total}}\left( \{\gamma_i\},\{\mu_i\} \right) = J\sum\limits_{i = 1}^N {\delta _i^{\left( \Delta  \right)} }  + I\sum\limits_{i = 1}^N {\left( {1 - \delta _i^{\left( \Delta  \right)} } \right) \sum\limits_{j = 1}^{2m} \delta \left( {\mu _i^j ,1} \right)}.  \hfill \\
\end{gathered}
\end{equation}

The first term,
\begin{equation}
\label{ham-basic}
-\beta H_{\text{polymer}}\left(\{\gamma_i\}\right)=J\sum\limits_{i=1}^{N}\delta _{i}^{(\Delta )}
\end{equation}
is the Potts-like interaction between different parts of the polymer. Here $\beta=T^{-1}$ is inverse temperature, $N$ the number of repeated units, and $J=U/T$ the temperature-reduced energy of hydrogen bonding. The short-hand notation is exploited, eg.
\begin{equation}
\delta_{j}^{(\Delta )}=\prod_{k=0}^{\Delta-1}\delta (\gamma_{j-k},1),
\label{deltaprod}
\end{equation}
where $\delta (x,1)$ stands for the Kronecker symbol and  $\gamma_{l}=1,\ldots,Q$ is the spin variable, which can be interpreted as being the number of rotating isomeric states of each repeated unit, with values between 1 and $Q$. The case when $\gamma_{l}$ = 1 denotes the helical state whereas all other ($Q-1$) cases correspond to coil states. $Q$ is the number of conformations of each repeated unit and thus describes the conformational variability. The Kronecker delta inside the Hamiltonian ensures the energy $J$ emerges only when all $\Delta$ successive neighboring repeated units are in the helical conformation. Thus, restrictions on backbone chain conformations imposed by hydrogen bond formation are indirectly taken into account. Estimates on the structure parameters for polypeptides were $\Delta =3$ and $Q=60-90$ \cite{biopoly1,biopoly2} and for double-stranded homo-polymeric DNA, $\Delta =10-15$ and $Q=3-5$  \cite{physa}. This GMPC model has been shown to be closely related to the Waiko-Sato-Eaton-Munoz (WSEM) model (compare the Eq.[10] of \cite{bad10} and Eq.[1] of \cite{wsme1} or \cite{wsme2}).
The second term,
\begin{equation}
\label{hamcs}
-\beta H_{\text{solvent}} = I\sum\limits_{i = 1}^N {\left( {1 - \delta _i^{\left( \Delta  \right)} } \right) \sum\limits_{j = 1}^{2m} \delta \left( {\mu _i^j ,1} \right)},
\end{equation}
represents the explicit interaction with the solvent, where $I=\frac{U_{ps}}{T}$ is the reduced energy of a polymer-solvent H-bond. 
Due to the presence of the term $1-\delta _i^{(\Delta)}$ in Eq.~(\ref{hamcs}), as opposed to the $\delta _i^{(\Delta)}$ term in Eq.(\ref{ham-basic}), the solvent is competing with the polymer for H-bond formation, depending on the ratio $J/I$.
As shown below, this competition gives rise to some interesting behavior.

A few remarks are in order regarding descriptions of solvents used here \cite{biopoly1,bad8}.  They are classified according to their relatively high or low molecular weight and according to their types of binding, i.e. either reversible or irreversible. Additionally, reversibly binding solvents can be divided into major groups according to their mechanism of interaction with the biopolymer. That is, those that compete for hydrogen bond formation with repeated units of the biopolymer (competing solvent) and those that do not (non-competing solvent). For simplicity in the present study only the low molecular weight competing solvent is considered. 

It is assumed (i) Polymer-solvent interactions depend on the state (orientation) of solvent molecules with respect to the repeated unit, and there are $q$ possible discrete orientations of each solvent molecule; (ii) A spin variable $\mu_i$, with values from $1$ to $q$ is assigned to each repeated unit $i$. Orientation number $1$ is the bonded one, with energy $E$.

Some solvents, such as water and urea are able to form hydrogen bonds with nitrogen basis of DNA or peptide groups of amino acids in proteins \cite{molbiol,gros,cantor,war,she}. We assume repeated units that are not bonded by intra-molecular H-bonds, to be free to form polymer-solvent intermolecular bonds. When one intra-molecular H-bond is broken, two binding sites for a solvent molecule become vacant. Thus, in the case of polypeptides there are only two binding sites per repeated unit, while in the case of DNA there are four ($2 \times 2$ for A-T pair) or six ($3 \times 2$ for G-C pair) binding sites, so $2m$ $(m=1,2,3...)$ spin variables are required to describe the interaction between solvent molecules and each repeated unit. The reduced energy $J$ of the Hamiltonian in Eq.~(\ref{ham-basic}) now becomes $J=m\frac{(U_{pp}+U_{ss})}{T}$, where $U_{pp}$ and $U_{ss}$ are the energies of polymer-polymer and solvent-solvent H-bonds, respectively. 

Using the simple identity,
\begin{eqnarray}
\label{identity}
\exp\left(J \delta_i^{(\Delta)}\right) &=& 1+\left(e^{J}-1\right) \delta_i^{\left(\Delta\right)},
\end{eqnarray}
the partition function associated with Eq.~(\ref{hamtotal}) can be quickly recast in the following form, 
\begin{equation}
\label{partfunctot}
\begin{gathered}
Z = \sum\limits_{\left\{ {\gamma _i } \right\},\left\{ {\mu _i } \right\}} 
{ \exp \left( - \beta H_{\text{total}}(\{\gamma _i\},\{\mu _i\})\right)}
=\hfill \\ \sum\limits_{\left\{ {\gamma _i } \right\}} \prod\limits_{i = 1}^N  [ {1+V\delta _i^{\left( \Delta  \right)}} ] \sum\limits_{\mu_i^1 =1}^q \sum\limits_{\mu_i^2 =1}^q ... \sum\limits_{\mu_i^{2m} =1}^q \prod\limits_{j = 1}^{2m}  \left[ {1+R(1-\delta _i^{\left( \Delta  \right)}}) \delta \left( {\mu _i^j ,1} \right) \right].
\end{gathered}
\end{equation}
\noindent Here $V=e^{J}-1$ and $R=e^{I}-1$.

As detailed in the Appendix, solvent degrees of freedom can then be traced out analytically to obtain 
\begin{equation}
\label{partfunctot3}
Z(\widetilde{V},R,Q,q,m)=(e^I+q-1)^{2mN}\times Z_0(\widetilde{V},Q).
\end{equation}
where
\begin{equation}
\label{Vtilde_discuss}
\widetilde{V}+1=\exp[\widetilde{J}]=\exp[\widetilde{U}/T]=
\left[{ \frac{e^{\frac{U_{pp}+U_{ss}}{T}}}{\left(1+\frac{e^{\frac{U_{ps}}{T}}-1}{q}\right)^2} }\right]^m=
\left[{ \frac{e^{\frac{1}{t}}}{1+\frac{e^{\frac{1+\alpha}{t}}-1}{q}} }\right]^{2m},
\end{equation}
and where 
\begin{equation}
\label{alpha}
\alpha=\frac{2U_{ps}-U_{pp}-U_{ss}}{U_{pp}+U_{ss}}
\end{equation}
is a parameter which sets the balance between polymer-polymer and polymer-solvent attraction. It proved convenient in Eq.(\ref{Vtilde_discuss}) to introduce the reduced temperature $t=2 T/(U_{pp}+U_{ss})$. 

Thus, the partition function of the original model with solvent can be reduced to the same model without solvent and renormalized interactions up to a multiplicative prefactor that depends on solvent properties only.

\section{Results and discussion}

\subsection{Inverse (reentrant) helix-coil transition arising from competition for hydrogen bond formation between water and polymer}

Our model incorporates the possibility of an inverse (reentrant) helix-coil transition whose origin stems mainly from competition between water and polymer for hydrogen bond formation, modeled through different values of $\alpha$. This is obvious directly from the Hamiltonian in Eq.~\ref{hamtotal}, as the two parts it is comprised of contain the same Kronecker $\delta$ symbol with opposite signs. As usual the Kronecker symbol is equal to one if an intra-molecular hydrogen bond exists, and zero otherwise. 

We note that the original Hamiltonian (\ref{hamtotal}) is an extension of the one used in Ref \onlinecite{biopoly1} in that both an arbitrary range, $\Delta$, of the intra-polymer interactions and an arbitrary number, $m$, of solvent states have been used, as opposed to $\Delta=3$ and $m=2$ fixed in Ref \onlinecite{biopoly1}. The same analysis based on the transfer-matrix formalism described in Ref \onlinecite{biopoly1} can then be carried out. Only final results are reported here.
Two quantities are of particular interest. First, the helicity degree, the average fraction of hydrogen bonded repeated units, 
\begin{equation}
\theta=\langle\delta _{i}^{(\Delta)}\rangle=\frac{1}{N}\frac{\partial }{\partial J} \ln Z;
\label{helicity}
\end{equation}
where $\langle \ldots \rangle$ is the usual thermal average over the original Hamiltonian (\ref{hamtotal}). 
The correlation length, corresponding to the spatial scale of correlation damping along the chain at large distances, can also be 
readily computed, 
\begin{equation}
\xi=\ln^{-1}\left(\frac{\lambda_1}{\lambda_2}\right);
\label{ksi}
\end{equation}
where $\lambda_1$ and $\lambda_2$ are the first and second leading eigenvalues of the transfer matrix (see Ref.\onlinecite{biopoly1} for details).

\begin{figure}[!ht]
\begin{center}
\includegraphics[width=9cm]{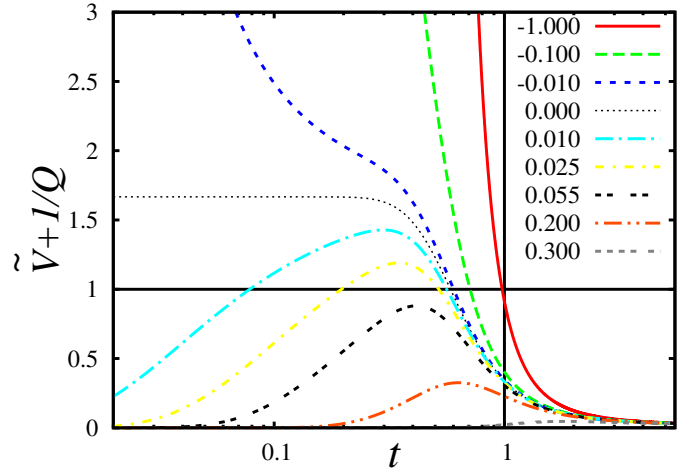}
\caption{\label{fig1} (Color online) Dependence of the stability parameter $(\widetilde{V}+1)/Q$ on reduced temperature $t=2 T/(U_{pp}+U_{ss})$ for the polypeptide (PP) parameter set: $\Delta=3$, $Q=60$ with $q=10$, $m=1$. Curves at different values of $\alpha=\frac{2U_{ps}-U_{pp}-U_{ss}}{U_{pp}+U_{ss}}$ are shown. Note, the point $(1,1)$ corresponds to the solvent-free transition point.}
\end{center}
\end{figure}

The helix-coil transition occurs whenever $\xi$ becomes large (see Eq.(\ref{ksi})) and hence when $\lambda_1\approx\lambda_2$. 
On the other hand it was previously shown \cite{biopoly1,bad10} that this corresponds to the condition $\widetilde{V}+1\approx Q$.
Note, in the absence of solvent, with $\Delta=2$ this is identical to the condition of vanishing $\Delta G$.  This is the change in Gibbs free energy between the helix and the coil state, just as obtained from the Zimm-Bragg model \cite{polsher} for $s=e^{-\Delta G/T}=1$, in terms of an entropy-enthalpy compensation.

Figure~\ref{fig1} depicts the stability parameter $(\widetilde{V}+1)/Q$ as a function of the reduced temperature $t=2 T/(U_{pp}+U_{ss})$ for several values of the parameter $\alpha$ ranging from $\alpha=-1$, corresponding to the absence of solvent, to $\alpha=0.3$, where the solvent plays a significant competitive role. As can be seen from the elementary analysis of Eq.~\ref{Vtilde_discuss} as shown in Fig 1, the $(\widetilde{V}+1)/Q$ curve is monotonic at $\alpha$ values from -1 to 0. At exactly $\alpha=0$ the curve has a plateau at low temperatures, while at $\alpha > 0$ a maximum appears. This maximum becomes lower with increased $\alpha$. Therefore in the range of $-1 < \alpha < 0$, the curve has only one intercept with the $f(t)=1$ line at temperatures close to $t=1$, indicating a regular helix-coil transition. When $\alpha > 0$ the situation is slightly more complicated, and either none or two intercepts exist corresponding to the reentrant transition at low temperature and the normal helix-coil transition at high temperatures. This happens because at first, as $\alpha$ increases above zero, the water-polymer energy $2 U_{ps}$ becomes slightly larger than $U_{pp}+U_{ss}$ and there is a competition between polymer-polymer and polymer-water hydrogen bond formation. In terms of the Gibbs free energy of the transition, existence of two intercepts of the stability parameter $(\widetilde{V}+1)/Q$ with $1$, would mean that $\Delta G$ is no longer monotonic with temperature, and entropy-enthalpy compensation happens in a more complex, non-linear way. In Fig.~\ref{fig1} there exists a well defined value of $\alpha \approx 0.04$ above which $(\widetilde{V}+1)/Q<1$ at all temperatures, making the intercept impossible\footnote{The limiting value of $\alpha$ is of course different at other $Q,q,m$ values, however, qualitative picture is the same.}.  This happens because the polymer-solvent energy $2 U_{ps}$ becomes so high compared to $U_{pp}+U_{ss}$, that mainly intermolecular hydrogen bonds are formed and the polymer preferably remains in the coil state. There still can be a certain amount of intra-molecular hydrogen bonding.  However, as shown below the helicity degree is always below one in this regime. 

Eq.~\ref{Vtilde_discuss}, visualised in Fig.~\ref{fig1} represents the result of tracing out the solvent degrees of freedom and sets the microscopic basis of the temperature dependence of the transition free energy. In the language of the Gibbs free energy, this means that $\Delta G$ is not a quadratic function of temperature as typically assumed within the Hawley framework \cite{hawley} (also in Refs.~\onlinecite{privalov,smeller,tanaka,graziano}).  Instead, it is a curve with maximum when $\alpha > 0$ and monotonically decreases otherwise. 

\begin{figure}[ht]\begin{center}
\includegraphics[width=8cm]{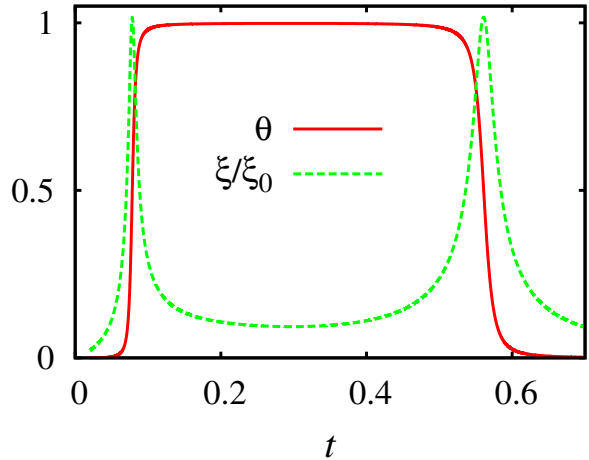}
\caption{\label{fig2} (Color online) Helicity degree $\theta$ and spatial correlation length $\xi$ plotted \emph{vs} temperature. Reduced units were used with the same parameter set as in Fig.~\ref{fig1}. $\alpha$ = $0.01$.}
\end{center}\end{figure}

Additional insight into origins of the transitions can be achieved by considering the helicity degree, $\theta$, and the correlation length, $\xi$. As clearly shown in Fig.\ref{fig2}, at $\alpha=0.01$ $\theta$ is zero at very low temperatures, suddenly increases to $1$ remaining at this plateau up to temperatures of order $1$ (in reduced units), and then drops back to $0$ at higher temperatures.
There are also two peaks of $\xi$ and the temperature values where $\theta=1/2$ and peak positions of $\xi$ are comparable. These transition points coincide with those determined from the intercepts with the $f(t)=1$ function in Fig.~\ref{fig1} (see the curve corresponding to $\alpha=0.01$). These quantities unambiguously indicate the presence of a regular helix-to-coil transition at high temperatures and an additional reentrant helix-to-coil transition at low temperatures.

Note that the maxima of correlation lengths for both high and low temperature transitions are equal. This is readily understood in terms of the cooperativity of the GMPC model \cite{biopoly2} since $Q$ and $\Delta$ are not altered upon a change in $\alpha$. 

\subsection{Reentrance in polynucleotides}

While the phenomenon of the reentrant protein folding (cold denaturation) transition has long been considered in protein melting, much less is known both experimentally and theoretically about the reentrant transition in DNA. 

In a previous study \cite{physa} we showed how the model with the Hamiltonian in Eq.~\ref{ham-basic} can be applied to describe the helix-coil transition in DNA provided that long-range loop formation is neglected. Thus, the considerations described above are also applicable to DNA and the existence of reentrant denaturation appears rather naturally since hydrogen bonds play an important role in DNA as well. This is confirmed by a number of experimental observations and seems quite obvious since water is the natural solvent for both.

Dubins \emph{et al} reported on pressure effects in double stranded nucleic acid melting \cite{dubins}. With the help of Hawley's phenomenological theory \cite{hawley_theory} they showed that there is a maximum point of the pressure-temperature diagram around 50 $^{\circ}$C where the nucleic acid is destabilized by pressure at temperatures lower than that and stabilized at higher temperatures. Using optical absorbance, they reported that the DNA/RNA hybrid duplex, poly(dA)poly-(rU) in 20 mM NaCl undergoes a pressure-induced helix-to-coil transition at room temperature under elevated pressure. Rayan and Macgregor \cite{macgregor}
reported the spectrophotometric observation of destabilization for poly[d(A-T)] and poly[d(I-C)] at increased pressure and various co-solvent concentrations.
Thus, there is considerable experimental and theoretical evidence for the existence of a reentrant helix-coil transition in polynucleotides.

\subsection{Phase diagram of the helix-coil transition and the meaning of parameter $\alpha$}

\begin{figure}[ht]\begin{center}
\includegraphics[width=8cm]{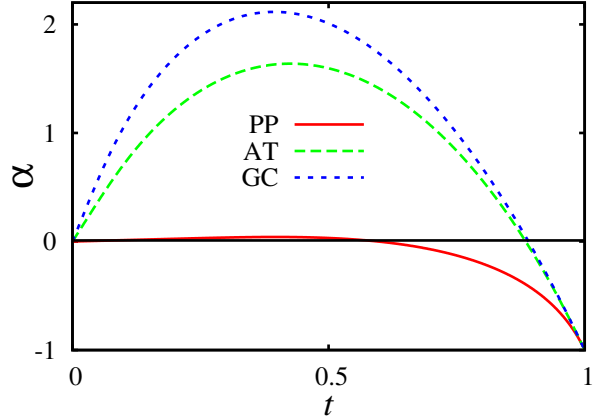}
\caption{\label{fig3} (Color online) Phase diagrams in variables $\alpha$ and $t$ for the polypeptide (PP) and two homopolymeric DNA's: AT and GC. For the PP case the variable set was the same as that in Fig.~\ref{fig1}. DNA cases have $\Delta=10$, $Q=3$, $q=10$; $m=2$ for AT and $m=3$ for GC. The regions below the curves correspond to the helix state, while regions above are the coil state. The straight line corresponds to the case shown in Fig.~\ref{fig2} and is the $\alpha=0.01$ curve. There are two intercepts with the PP curve, indicating a low-temperature reentrant transition together with the regular helix-to-coil transition at higher temperatures.}
\end{center}\end{figure}

As remarked in connection with Fig.~\ref{fig1}, the helical state is not possible for all values of temperature and $\alpha$. Fig.~\ref{fig3} illustrates this point in the $\alpha$ vs reduced temperature plane, where all points above (below) the depicted $\alpha (t)$ curves are in the coil (helix) state. The cases of the polypeptide chain (PP), as well as regular DNA heteropolymers of the AT and GC type have been considered. Note, the phase diagrams for DNA are much sharper than the polypeptide curve. Sharpness of these curves is associated with the higher cooperativity of the helix-coil transition of DNA represented by higher values of $\Delta$ (reflecting more rigidity) and by smaller $Q$, corresponding to strong reduction of available conformational space compared to the PP case. 

The main obstacle in comparing results presented here with experiments hinges on the general difficulty of defining a pressure in spin models. This can however be accounted for in the following indirect way. Consider the parameter 
\begin{equation}
\frac{\alpha}{t} = \frac{2U_{ps}-(U_{pp}+U_{ss})}{2 T},
\label{FH}
\end{equation}
\noindent where $\alpha=\frac{2U_{ps}-(U_{pp}-U_{ss})}{U_{pp}+U_{ss}}$ plays here a role similar to the Flory-Huggins $\chi$ parameter in polymer theory \cite{gros,Rubinstein03}. In addition here $\alpha$ is a measure of the relative influence of the polymer-polymer attractions ($\alpha <0$) versus polymer-solvent attractions ($\alpha >0$). \par

Usually in order to avoid effects of inter-polymer interactions helix-coil melting experiments are performed on dilute solutions of biopolymers. Since the volume fraction of polymer is extremely small at low polymer concentrations, effects of hydrostatic pressure on polymer conformations can only be indirect occuring through changing water properties.  It has been shown \cite{FH_pressure} that hydrostatic compression loosens water hydrogen-bond structure near the monomer unit. Thus, increasing hydrostatic pressure effectively promotes transfer of water molecules from the hydrogen bonded cluster to non-structured solvent with subsequent binding to the macromolecule. Therefore increasing hydrostatic pressure makes binding of water molecules with the polymer more thermodynamically favorable than binding to other water molecules, resulting in decreased values of $U_{ss}$. To summarize, it can concluded that increased pressure corresponds to increased values of $\alpha$ and \emph{vice versa}. 

Native, ordered phases are usually found in regions where both pressures and temperatures are low so that the energies of polymer-polymer and solvent-solvent interactions in this region overwhelm the polymer-solvent interaction energies and $\alpha < 0$. Conversely, at low temperature an increase of pressure would result in a disordered coil phase where $\alpha > 0$.

This line of logic in view of Fig.~\ref{fig3} suggests the reentrant coil-helix transition we observed is the counterpart of high pressure denaturation observed experimentally. In general, stability of the polymer-water system can be altered in many ways.  These include increased pressure \cite{hawley_theory,hawley,smeller,tanaka,macgregor}, addition of denaturants \cite{hopkins,urea}, changes in pH \cite{privalov,kauzmann}, application of osmotic pressure \cite{chris} or stretching force \cite{rouzina,hanke}. Since all these factors can alter the balance between inter and intra-molecular hydrogen bonds, they can be modeled by changes in $\alpha$, which in its turn mirrors changes of the Flory-Huggins parameters. 

\subsection{Phase diagram of DNA under pulling and unzipping and the inversion of force effect at low and high forces}

While there is a long history of pressure, temperature induced melting in biopolymers, the role of pulling force as an external parameter is relatively new. Long ago Hawley and Macleod \cite{hawley} studied effects of pressure on the melting temperature of Clostridium perfringens DNA. Unlike the heat induced unfolding temperature of proteins, the melting temperature of their DNA did not display any curvature, and was instead a purely linear function of pressure over the range studied (0-400 MPa). The first numerical prediction of the force-induced reentrant helix-coil transition dates back to 1993 \cite{prohofsky}. Since then a number of other studies have supported this prediction in DNA. Marenduzzo \emph{et al} \cite{marenduzzo} and Orlandini \emph{et al} \cite{orlandini} predicted existence of unzipping using a computational model. Phase diagrams for pulling have been reported by Rudnick and Kuriabova \cite{rudnick} and Rahi \emph{et al} \cite{rahi}. Experimental observations of an inversion of force effects on DNA stability (from stabilizing at forces $<$ 7 pN, to destabilizing at higher forces) were reported [\onlinecite{rouzina}] and the corresponding phase diagram for pulling was shown in \cite{williams}. A similar phase diagram was reported in the theoretical work by Hanke \emph{et al} \cite{hanke} in (force, temperature) variables. The phase diagrams in \cite{williams} (experiment, see Fig.5 of \cite{williams}) and \cite{hanke} (theory, see Fig. 2a of \cite{hanke}, case where $A=0.01$) are similar, in that at small forces there are two regions where melting temperature increases with force. In these regions force has a stabilizing effect on DNA. 

Our Fig.~\ref{fig3} reproduces the main qualitative features of the phase diagram for DNA melting (see, e.g. Fig. 4 of \cite{williams}) and agrees well with Refs.~\cite{marenduzzo,orlandini,rudnick,rahi}). However, contrary to those studies we did not observe an increase in stability with increasing $\alpha$ on the right side of the phase diagram. This might be due to specific features of DNA associated with large loops which were omited in our studies \cite{physa}. Another possible source of this discrepancy could be the polyelectrolyte nature of polynucleotides. For instance when water-polymer hydrogen bonds are broken and water molecules return to the bulk, helix formation is then hindered by electrostatic repulsion between negatively charged nearest neighbors that were previously screened by water. 

\subsection{Pressure versus force: what is the difference?}

Our analysis provides an explanation as to why it is much more difficult to observe the reentrant transition under high  pressure compared to experiments using a pulling force. As previously mentioned, hydrostatic pressure can only indrectly affect the balance between polymer-solvent and polymer-polymer bonds. A very large variation of pressure is required to induce conformational changes in dilute solutions of biopolymers. On the other hand, applied force operates directly on $U_{ps}$ and $U_{pp}$, meaning small changes in pulling force induce large changes in $\alpha$. This because it is much harder to compress several milliliters of water enough to produce significant changes in biopolymers occupying a relatively small volume of the solution, compared to pulling a single molecule. 

Further going back to the phase diagrams in Fig.~\ref{fig3}, in the case of the polypeptide, small changes in $\alpha$ shift the reentrant transition point to moderate temperatures making experimental observation of cold denaturation possible. In contrast on phase diagrams for DNA small increments of $\alpha$ do not result in significant changes of stability and a much wider range of experimental pressures is required to induce the transition. As opposed to pressure, pulling more strongly affects the hydrogen bonding equilibrium in DNA allowing observation of the coil state at experimentally feasible temperatures. 

Together these facts explain why the reentrant transition in DNA is very difficult to experimentally observe in pressure-based experiments, but is readily observed in experiments using a pulling force.

\subsection{Buhot and Halperin approach to helix-coil transition in polypeptides under stretching}

Recently, Buhot and Halperin offered an elegant study of extension behavior of helicogenic polypeptides. They have explicitly added extension force into the Zimm and Bragg description. In one of their papers a step-like behavior of helicity degree versus normalised force has been reported \cite{buhot1}, without any model for solvent. A thorough study of Tamashiro and Pincus reported no step-like behavoir of helicity degree within the same model \cite{tamashiro}. Later, Buhot and Halperin mention that the appearance of a step on helicity degree is an artifact of the approximations they have used \cite{buhot2}. The theories by Buhot and Halperin and Tamashiro and Pincus are in good agreement with some experimental studies \cite{buhotexper}, however there was no intention to study the effect of stretching onto reentrant helix-coil transition, and no solvent effects were included into the consideration. 

It could be interesting to experimentally study the pulling effects in the region of temperatures where the reentrant coil-helix transition takes place. Such a study would justify the necessity of solvent model inclusion into the approach offered by Buhot and Halperin.
\section{Conclusions}
\label{sec:conclusions}
In this paper within the framework of the Generalized Model of Polypeptide Chains (GMPC) the possibility of a reentrant helix-coil transition at low temperatures has been investigated. The GMPC is a Potts-like spin model where competition between the tendency to form a helix state, as enforced by polymer-polymer interactions $J$, is balanced by the configurational entropy as measured by the remaining $Q-1$ coil states. In the presence of explicit solvent interactions the equilibrium is modified by adding a competing polymer-solvent H-bond formation (strength $I$), and the solvent configurational entropy measured by the remaining $q-1$ states. The number of solvent spin variables depends whether a polypeptide or a DNA is considered, and in our model is considered an arbitrary variable $2m$ in the model, so that both cases can be treated within the same unified model.

We have shown how solvent degrees of freedom can be traced out exactly to obtain an effective GMPC model with renormalized interactions. This is then studied following the recipes outlined in past work in the absence of solvent interactions. A low-temperature coil-helix reentrant transiton is found in terms of a parameter akin to the Flory-Huggins parameter. This is indicated by the appearence of a sudden drop in the helicity, $\theta$, and confirmed by a second low-temperatures peak in the correlation length, $\xi$. A global phase diagram separating helical and coil states has also been illustrated.

It should be emphasised that the key to our success in qualitatively describing the abovementioned systems and experimental situations stems from the possibility of modeling the most relevant solvent features with respect to the helix-coil transition, within a sufficiently simple theoretical scheme that allows analytical treatment. 
  
\begin{acknowledgments}
AB and AG acknowledge the support from PRIN-COFIN2007B58EAB grant.
\end{acknowledgments}

\appendix
\section{Exact integration of the solvent degrees of freedom}
In Eq.~(\ref{partfunctot}) we have
\begin{equation}
\label{L_i}
\begin{gathered}
L_{\text{comp.solv.}}({\left\{ {\gamma _i } \right\}}) \equiv \sum\limits_{\left\{ {\mu_i^j } \right\}} \prod\limits_{j = 1}^{2m} \left[ {1+R(1-\delta _i^{\left( \Delta  \right)}}) \delta \left( {\mu _i^j ,1} \right) \right]= \\ \sum\limits_{\mu_i^1 =1}^q \sum\limits_{\mu_i^2 =1}^q ... \sum\limits_{\mu_i^{2m} =1}^q  \{ 1+R(1-\delta _i^{\left( \Delta  \right)}) \sum\limits_{j=1}^{2m} \delta \left( {\mu _i^j ,1} \right)+ R^2(1-\delta _i^{\left( \Delta  \right)}) \sum\limits_{j<k}\delta \left( {\mu _i^j ,1} \right)\delta \left( {\mu _i^k ,1} \right)+
 \\ R^3(1-\delta _i^{\left( \Delta  \right)}) \sum\limits_{j<k<l}\delta \left( {\mu _i^j ,1} \right)\delta \left( {\mu _i^k ,1} \right)\delta \left( {\mu _i^l ,1} \right)+... \\ R^{2m}(1-\delta _i^{\left( \Delta  \right)})\delta \left( {\mu _i^1 ,1} \right) \delta \left( {\mu _i^2 ,1} \right)... \delta \left( {\mu _i^{2m} ,1} \right) \} = \\ q^{2m}+(1-\delta _i^{\left( \Delta  \right)})\left[2mRq^{2m-1}+C_{2m}^2 R^2 q^{2m-2}+C_{2m}^3 R^3 q^{2m-3}+...+R^{2m}\right]= \\q^{2m}+(1-\delta _i^{\left( \Delta  \right)})\left[q+R\right]^{2m}-(1-\delta _i^{\left( \Delta  \right)})q^{2m}= (q+R)^{2m} \left[ 1-\delta _i^{\left( \Delta  \right)}+\frac{q^{2m} \delta _i^{\left( \Delta  \right)}}{(q+R)^{2m}}\right].
\end{gathered}
\end{equation}
where $V=e^{J}-1$ and $R=e^{I}-1$ have been introduced in the main text, and where the binomial coefficients $C_{n}^{m}=n!/(m!(n-m)!)$ have been
exploited. 
\noindent The last expression can be inserted instead of the solvent part of Eq.~(\ref{partfunctot}), resulting in
\begin{equation}
\label{partfunctot2}
\begin{gathered}
Z = \sum\limits_{\left\{ {\gamma _i } \right\}} \prod\limits_{i = 1}^N \left[ {1+V\delta _i^{\left( \Delta  \right)}} \right] \times (q+R)^{2m} \left[ 1-\delta _i^{\left( \Delta  \right)}+\frac{q^{2m} \delta _i^{\left( \Delta  \right)}}{(q+R)^{2m}}\right ] =\\ (q+R)^{2mN} \sum\limits_{\left\{ {\gamma _i } \right\}} \prod\limits_{i = 1}^N \left[ {1+\widetilde{V}\delta _i^{\left( \Delta  \right)}} \right] ,
\end{gathered}
\end{equation}

\noindent where
\begin{equation}
\label{Vtilde}
\widetilde{V}+1=\exp{(\widetilde{J})}=\frac{(V+1)q^{2m}}{(q+R)^{2m}}=\frac{\exp{(m\frac{(U_{pp}+U_{ss})}{T})} q^{2m}}{\left[ q-1+\exp{(\frac{U_{ps}}{T})}\right]^{2m}},
\end{equation}
\noindent
Then Eq.(\ref{partfunctot3}) is obtained.

\bibliographystyle{apsrev}
\bibliographystyle{apsrev}

\end{document}